\documentclass{jfm}

\usepackage{subfigure}
\usepackage{graphicx}
\usepackage{natbib}
\usepackage{bm}

\ifCUPmtlplainloaded \else
  \checkfont{eurm10}
  \iffontfound
    \IfFileExists{upmath.sty}
      {\typeout{^^JFound AMS Euler Roman fonts on the system,
                   using the 'upmath' package.^^J}%
       \usepackage{upmath}}
      {\typeout{^^JFound AMS Euler Roman fonts on the system, but you
                   dont seem to have the}%
       \typeout{'upmath' package installed. JFM.cls can take advantage
                 of these fonts,^^Jif you use 'upmath' package.^^J}%
      }
  \else
  \fi
\fi

\ifCUPmtlplainloaded \else
  \checkfont{msam10}
  \iffontfound
    \IfFileExists{amssymb.sty}
      {\typeout{^^JFound AMS Symbol fonts on the system, using the
                'amssymb' package.^^J}%
       \usepackage{amssymb}%

      }{}
  \fi
\fi

\ifCUPmtlplainloaded \else
  \IfFileExists{amsbsy.sty}
    {\typeout{^^JFound the 'amsbsy' package on the system, using it.^^J}%
     \usepackage{amsbsy}}
    {\providecommand\boldsymbol[1]{\mbox{\boldmath $##1$}}}
\fi

\newsavebox{\astrutbox}
\sbox{\astrutbox}{\rule[-5pt]{0pt}{20pt}}

\title[Velocity Space-Time Correlations in Turbulent Rayleigh-B\'{e}nard Convection]{Experimental Investigation of Longitudinal Space-Time Correlations of the Velocity Field in Turbulent Rayleigh-B\'{e}nard Convection}
\author[Zhou \emph{et al.}]{Quan ZHOU \thanks{Email address for correspondence: qzhou@shu.edu.cn}, Chun-Mei LI, Zhi-Ming LU \and Yu-Lu LIU}
\affiliation{Shanghai Key Laboratory of Mechanics in Energy and Environment Engineering, Shanghai Institute of Applied Mathematics and Mechanics, E-Institutes of Shanghai Universities, Shanghai University, Shanghai 200072, China}

\date{?? and in revised form ??}

\begin{document}

\maketitle

\begin{abstract}
We report an experimental investigation of the longitudinal space-time cross-correlation function of the velocity field, $C(r,\tau)$, in a cylindrical turbulent Rayleigh-B\'{e}nard convection cell using the particle image velocimetry (PIV) technique. We show that while the Taylor's frozen-flow hypothesis does not hold in turbulent thermal convection, the recent elliptic model advanced for turbulent shear flows \cite[]{he06pre} is valid for the present velocity field for all over the cell, i.e., the isocorrelation contours of the measured $C(r,\tau)$ have a shape of elliptical curves and hence $C(r,\tau)$ can be related to $C(r_E,0)$ via $r_E^2=(r-U\tau)^2+V^2\tau^2$ with $U$ and $V$ being two characteristic velocities. We further show that the fitted $U$ is proportional to the mean velocity of the flow, but the values of $V$ are larger than the theoretical predictions. Specifically, we focus on two representative regions in the cell: the region near the cell sidewall and the cell's central region. It is found that $U$ and $V$ are approximately the same near the sidewall, while $U\simeq0$ at cell center.
\end{abstract}


\section{Introduction}

Turbulent flows contain eddies with various scales. Turbulent kinetic energy is transferred from eddies with the largest scale of turbulence, $L$, at which energy is injected into the turbulence system, to eddies with the smallest scale of turbulence, $\eta$, at which energy is dissipated by fluid viscosity. Such cascade processes are usually characterized by the velocity structure functions, $S_p(r)=\langle|\delta_r v|^p\rangle$, defined as moments of velocity increments over a space separation $r$, where $\langle \cdots \rangle$ denotes a time average. Since the pioneering work of \cite{K41}, various theories and models have been put forwards to predict the scaling behaviors of the velocity structure functions in the so-called inertial range $\eta<<r<<L$ \cite[see, for reviews,][]{frisch1995, sreenivasan97arfm}. From experimental aspects, velocity measurements are usually carried out at a single fixed location, based on which time series of fluctuating velocities are obtained and the velocity structure functions, $S_p(\tau)=\langle|\delta_{\tau} v|^p\rangle$, are calculated as moments of velocity increments over a time separation $\tau$. To relate the properties of the experimentally measured $S_p(\tau)$ in time domain to theoretical predictions of $S_p(r)$ in space domain, one needs to invoke the Taylor's frozen-flow hypothesis \cite[]{T38}. The validity of such hypothesis demands low turbulent intensity and weak shear rates \cite[]{lumley96}. However, these conditions are not always met by actual flows of interest \cite[]{pinton1994jp}.

Another quantity that can also be used to characterize the cascade processes is the velocity space-time correlation function, defined as
\begin{equation}
C(r,\tau)=\frac{\langle v(\textbf{\emph{x}}+\textbf{\emph{r}},t+\tau)v(\textbf{\emph{x}},t)\rangle}{v_{rms}(\textbf{\emph{x}})v_{rms}(\textbf{\emph{x}}+\textbf{\emph{r}})},
\end{equation}
where $v(\textbf{\emph{x}})$ is one of the components of the velocity vector at position $\textbf{\emph{x}}$ and $v_{rms}(\textbf{\emph{x}})$ is the root-mean-square (r.m.s.) velocity at $\textbf{\emph{x}}$. For simplicity, we consider here only the situation that $\textbf{\emph{r}}$ is in the direction of $v(\textbf{\emph{x}})$, i.e. the longitudinal velocity correlations. [Studies of other kinds of velocity correlations, such as those between the wall-normal velocity component and the streamwise component, could be found in \cite{schumacher2006jfm}.] Again, the velocity temporal auto-correlation function $C(0,\tau)$, based on the pointwise measurements, is the quantity most often studied in experiments and Taylor's frozen-flow hypothesis is usually used to translate $C(0,\tau)$ in time domain to $C(r,0)$ in space domain. When Taylor's hypothesis holds, we have $C(r,\tau)=C(r_T,0)$ with
\begin{equation}
\label{eq:rt}
r_T=r-U_0\tau,
\end{equation}
where $U_0$ is the mean velocity of the flow. Such a relation implies that $C(r,\tau)$ would keep constant at increasing $r$ and $\tau$ once the value of $r-U_0\tau$ remains constant, which violates the basic properties of the correlations that $C(r,\tau)$ decays to zero at sufficient large separations. Therefore, Taylor hypothesis holds only for the limited ranges of $r$ and $\tau$.

Recently, based on a second order approximation to $C(r,\tau)$, \cite{he06pre} advanced an elliptic model for turbulent shear flows and proposed that $C(r,\tau)$ could be related to $C(r_E,0)$ via
\begin{equation}
\label{eq:re}
r_E^2=(r-U\tau)^2+V^2\tau^2,
\end{equation}
where $U$ is a characteristic convection velocity proportional to the mean velocity $U_0$ and $V$ is a characteristic velocity associated with the r.m.s. velocity and the shear-induced velocity \cite[]{he09pre}. Specifically, when $V$ vanishes, (\ref{eq:re}) is degenerated to the Taylor's hypothesis (\ref{eq:rt}), while Kraichnan's sweeping-velocity hypothesis \cite[]{kraichnan64} is obtained if $U$ vanishes. The He's elliptic model provides a useful tool for a large set of flow systems where the Taylor's hypothesis does not hold and hence the validity of the model in various practical flows needs to be tested.

In this paper, we want to validate the elliptic model in turbulent convection, an important class of turbulent flows that play an central role in many natural and engineering processes. The flow at hand is turbulent Rayleigh-B\'{e}nard (RB) convection, i.e. the convective motion of an enclosed fluid layer heated from below and cooled from above, which has received tremendous attention during the past few decades (Ahlers, Grossmann $\&$ Lohse 2009; Lohse $\&$ Xia 2010). Although it has long been recognized that the conditions for Taylor's hypothesis are often not met in the system \cite[see, e.g.,][]{shang2001pre}, single-point or time-domain measurements are employed by most experimental investigators for the studies of turbulent cascade processes of both the velocity and temperature fields (see the recent review paper, Lohse $\&$ Xia 2010, and references therein). To translate correctly the quantities measured in time domain to those in space domain, it is thus essential to validate the elliptic model in turbulent RB convection.

Recently, He, He $\&$ Tong (2010) verified indirectly the elliptic model via the local temperature data as in the bulk region of turbulent RB convection temperature behaves as a passive scalar (Calzavarini, Toschi $\&$ Tripiccione 2002; Xi \emph{et al.} 2009), which is driven by the velocity field via a linear equation. The authors showed that the elliptic relation (\ref{eq:re}) is valid for the temperature space-time correlations measured in the cell's sidewall region, but at the cell center they did not observe the predicted relation (\ref{eq:re}) for the temperature field. However, a passive additive may display characteristics so different from those of the advecting velocity field \cite[]{warhaft2000arfm}. In addition, the temperature field is nearly homogeneous in the cell's central region. The lack of temperature contrast would make the local temperature not follow the behaviors of the velocity field, e.g. the local velocity fluctuations show strong oscillations at the cell center \cite[]{qiu2004pof, zhou2009jfm}, while the oscillation is absent for temperature measured at the same location \cite[]{qiu2002pre}, and hence it is not surprising that (\ref{eq:re}) does not hold for temperature in the cell's central region. Therefore, it is highly desirable to verify the elliptic model \emph{directly} via the velocity field, which is the object of the present experimental investigation.

The remainder of this paper is organized as follows. We give detailed descriptions of the experimental apparatus and conditions and the measuring technique in $\S$2. Experimental results are presented and analyzed in $\S$3, which is divided into three parts. In $\S\S$3.1 and 3.2, we study the properties of longitudinal space-time correlation functions of the vertical velocity near the cell sidewall and at the cell center, respectively. Section 3.3 presents results of longitudinal space-time correlations of the vertical velocity along the cell's diameter at the middle height of the cell and of longitudinal space-time correlations of the horizontal velocity along the cell's central vertical axis. The properties of the characteristic velocities $U$ and $V$ are also investigated in detail in $\S$3.3. We summarize our findings and conclude in $\S$4.

\section{Experimental setup and parameters}

The convection cell is similar to that used in previous experiments \cite[]{sun2005jfm}, but has a different size \cite[]{zx2010prl_lds}. It is a vertical cylinder of height $H=50$ cm and inner diameter $D=50$ cm and hence of unit aspect ratio. Deionized and degassed water was used as the convecting fluid. The cell's sidewall is a plexiglas tube of 5 mm in wall thickness and a square-shaped jacket made of flat plexiglas plates and filled with water is fitted round the sidewall, which greatly reduced the distortion effect to the PIV images caused by the curvature of the cylindrical sidewall. The top and bottom plates are made of pure copper with nickel-plated fluid-contact surfaces. The thickness of the top plate is 3 cm and that of the bottom one is 1.5 cm. Four spiral channels of 1.2 cm in width and 1.5 cm in depth are machined into the top plate and the separation between adjacent channels is 1.1 cm. The channels start from the center and end near the edge of the plate. A silicon rubber sheet and a Plexiglas plate are fixed on the top to form the cover and also to prevent interflow between the channels. Each channel is connected to a separate refrigerated circulator (Polyscience 9712) that has a temperature stability of 0.01 $^{\circ}$C. The channels and the circulators are connected such that the incoming cooler fluid and the outgoing warmer fluid in adjacent channels always flow in opposite directions. Four quarter-circular Kapton film heaters, connected in parallel to a dc power supply (Xantrex XDC 300-20) with 99.99$\%$ long-term stability, are sandwiched to the back side of the bottom plate to provide constant and uniform heating. Therefore, the experiments were conducted under constant heating of the bottom plate while maintaining a constant temperature at the top plate. Eight thermistors are embedded beneath the fluid-contact surface of each conducting plate, equally spaced azimuthally at about one-third radius from the edge. The measured relative temperature differences among eight thermistors in the same plate are found to be smaller than 3$\%$ of that across the fluid layer.

During the experiment the entire cell was wrapped by several layers of Styrofoam and the cell was tilted by a small angle of about 0.5$^{\circ}$ (Ahlers, Brown $\&$ Nikolaenko 2006) so that the measurements were carried out within the vertical plane of the large-scale circulation. The mean temperature of water was kept at 29$^{\circ}$, corresponding to a Prandtl number $Pr=\nu/\kappa=5.5$. The experiment covered the range $5.9\times10^9\lesssim Ra\lesssim1.1\times10^{11}$ of the Rayleigh number $Ra=\alpha g\Delta T H^{3}/\nu\kappa$, with $g$ being the gravitational acceleration, $\Delta T$ the temperature difference across the fluid layer, and $\alpha$, $\nu$ and $\kappa$ being, respectively, the thermal expansion coefficient, the kinematic viscosity, and the thermal diffusivity of water. As all measurements give the same qualitative results, only that for $Ra=9.5\times10^{10}$ will be presented in this paper. At this $Ra$, the period of the large-scale circulation is around 62 sec.

\begin{figure}
\center
\resizebox{0.6\columnwidth}{!}{%
  \includegraphics{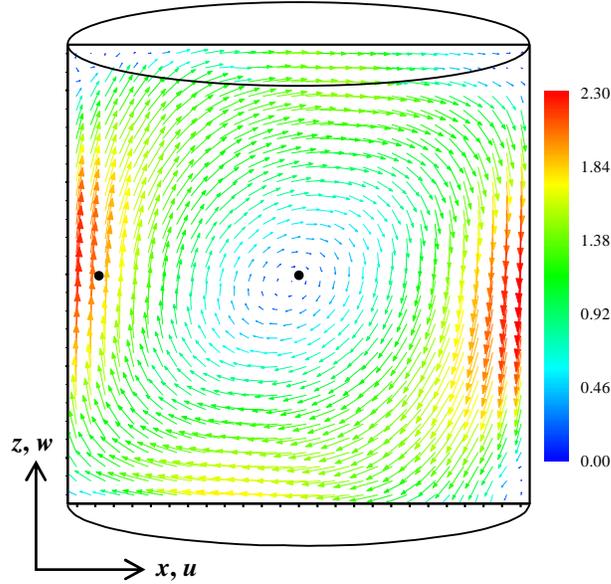}
} \caption{Sketch of the convection cell, the Cartesian coordinates used in the experiment, and the time-averaged vector map of the whole velocity field measured in the vertical plane of the large-scale circulation at $Ra=9.5\times10^{10}$ and $Pr=5.5$. For clarity, a coarse-grained vector map of size $32\times32$ is shown. The magnitude of the velocity $\sqrt{u_0^2+w_0^2}$ is coded in both color and the length of the arrows in units of cm/s. The time average is taken over a period of 145 min corresponding to 20 000 velocity frames. The two black dots in the figure mark the positions (x=-21.73cm, z=0cm) (see $\S$ \ref{sec:side}) and (x=0cm, z=0cm) (see $\S$ \ref{sec:center}), respectively.} \label{fig:fig1}
\end{figure}

\begin{figure}
\center
\resizebox{1\columnwidth}{!}{%
  \includegraphics{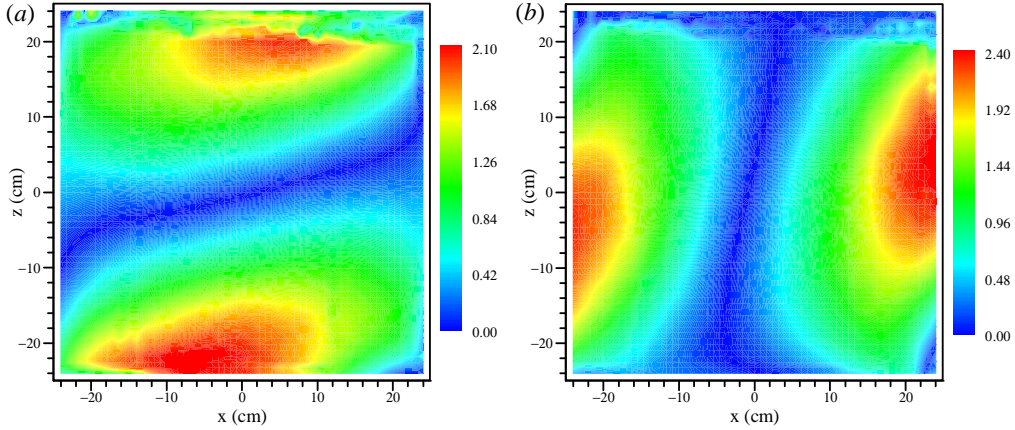}
} \caption{Color-coded contour maps of $|u_0/u_{rms}|$ (\emph{a}) and $|w_0/w_{rms}|$ (\emph{b}) obtained at $Ra=9.5\times10^{10}$ and $Pr=5.5$. Here, $u_0$ and $u_{rms}$ are the time-averaged and the r.m.s. horizontal velocities, respectively, and $w_0$ and $w_{rms}$ are the time-averaged and the r.m.s. vertical velocities, respectively.} \label{fig:fig2}
\end{figure}

The details of the particle image velocimetry (PIV) measurements in turbulent RB convection have been described and discussed by Xia, Sun $\&$ Zhou (2003) and Sun, Xia $\&$ Tong (2005), here we give only its main features. The laser lightsheet thickness is $\sim2$ mm and the seed particles are 50-$\mu$m-diameter polyamid spheres (density 1.03 g/cm$^3$). As the seed particles are neutrally buoyant, they are assumed to follow the motion of the fluid. The measuring region has an area of $49\times49$ cm$^2$ with a spatial resolution of 7.76 mm, corresponding to $63\times63$ velocity vectors. We chose the vertical rotation plane of the large-scale circulation as the laser-illuminated plane, defined as the $(x,z)$ plane. The Cartesian coordinate is defined such that the origin $(0,0)$ coincides with the cell center, the $x$ axis points to the right, and the $z$ axis points upwards. The horizontal velocity component $u(x,z)$ and the vertical one $w(x,z)$ were obtained. The experiment lasted 145 minutes in which a total of 20000 two-dimensional vector maps were acquired with a sampling rate of $\sim2.3$ Hz. Figure \ref{fig:fig1} shows the measured mean flow field, together with the Cartesian coordinates used in the experiment. In the figure, the magnitude of the mean velocity $\sqrt{u_0^2+w_0^2}$ are coded both by color and by the length of the arrows, where $u_0=\langle u(t)\rangle$ and $w_0=\langle w(t)\rangle$ are the time-averaged horizontal and vertical components of the velocity vector, respectively. One sees clearly that the mean flow is a clockwise rotatory motion with a relatively quiet central region and high velocity regions concentrated along the perimeter of the cell. To see whether the Taylor's frozen-flow hypothesis is valid or not in turbulent RB system, we plot in figure \ref{fig:fig2} the spatial distributions of the ratios between the time-averaged and the r.m.s. velocities, $|u_0/u_{rms}|$ [figure \ref{fig:fig2}(\emph{a})] and $|w_0/w_{rms}|$ [figure \ref{fig:fig2}(\emph{b})], where $u_{rms}=\sqrt{\langle[u(t)-u_0]^2\rangle}$ and $w_{rms}=\sqrt{\langle[w(t)-w_0]^2\rangle}$ are the r.m.s. horizontal and vertical velocities. If Taylor hypothesis holds, the values of $|u_0/u_{rms}|$ or $|w_0/w_{rms}|$ should be much larger than 1. In the figures, it is seen that both $|u_0/u_{rms}|$ and $|w_0/w_{rms}|$ are nearly zero in the cell's central region and $|u_0/u_{rms}|$ and $|w_0/w_{rms}|$ increase, respectively, in the vertical and horizontal directions. The maximization of $|u_0/u_{rms}|$ occurs near the top and bottom plates with a maximum value $|u_0/u_{rms}|_{max}\simeq2.1$, and $|w_0/w_{rms}|$ reaches its maximum value near the cell sidewall with $|w_0/w_{rms}|_{max}\simeq2.4$. For such large r.m.s. velocities, Taylor's frozen-flow hypothesis is then not expected to be valid.

The PIV technique provides us a convenient tool to directly and simultaneously measure the local fluctuating velocities at multi-points in a particular plane of interest. With the measured $u(x,z,t)$ and $w(x,z,t)$, one can obtain the longitudinal space-time cross-correlation functions for both the horizontal and vertical velocities, respectively, defined as
\begin{equation}
C_u(r,\tau; z)=\frac{\langle[u(x+r,z,t+\tau)-u_0(x+r,z)][u(x,z,t)-u_0(x,z)]\rangle}{u_{rms}(x+r,z)u_{rms}(x,z)}
\end{equation}
and
\begin{equation}
C_w(r,\tau; x)=\frac{\langle[w(x,z+r,t+\tau)-w_0(x,z+r)][w(x,z,t)-w_0(x,z)]\rangle}{w_{rms}(x,z+r)w_{rms}(x,z)}.
\end{equation}

\section{Results and discussions}

\subsection{Near the cell sidewall}
\label{sec:side}

\begin{figure}
\center
\resizebox{1\columnwidth}{!}{%
  \includegraphics{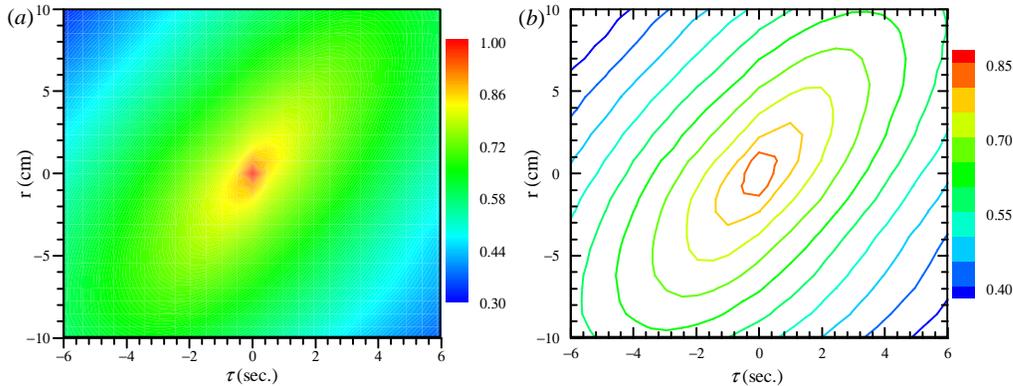}
} \caption{(\emph{a}) Space-time correlations $C_w(r,\tau; x)$ as a function of $r$ and $\tau$ measured near the cell sidewall ($x=-21.73$ cm and $z=0$ cm). Here, the amplitude of $C_w(r,\tau; x)$ is coded by color. (\emph{b}) The corresponding isocorrelation contours of $C_w(r,\tau;x)$ with the color-coded correlation amplitude varying from 0.4 to 0.85 at increments of 0.05 (outer to inner contours).} \label{fig:fig3}
\end{figure}

We first study the properties of longitudinal space-time correlations for the vertical velocity, $C_w(r,\tau; x)$, near the cell sidewall ($x=-21.73$ cm) at the middle height of the cell ($z=0$ cm), where the mean and the r.m.s velocities are of the same order [see figure \ref{fig:fig1}(\emph{b})]. Figure \ref{fig:fig3}(\emph{a}) shows the flood contours of the measured $C_w(r,\tau; x)$ as a function of separations $r$ and $\tau$ and the corresponding isocorrelation contours are plotted in figure \ref{fig:fig3}(\emph{b}). By definition, the maximization of $C_w(r,\tau; x)$ occurs at the origin with a maximum value of $C_w(0,0; x)=1$. As $C_w(r,\tau; x)$ decays fast near the origin, our present resolution could not resolve properly the isocorrelation contours when $C_w(r,\tau; x)\gtrsim0.85$. If the Taylor's hypothesis relation (\ref{eq:rt}) is valid for the present flow field, the isocorrelation contours of $C_w(r,\tau; x)$ should be straight lines. However, for the $r$- and $\tau$-range studied, one sees in figure \ref{fig:fig3}(b) that the isocorrelation contours of the measured $C_w(r,\tau;x)$ are the elongated and closed curves, rather than straight lines, and seem to have a shape of elliptical curves that can be described well by (\ref{eq:re}). Furthermore, all isocorrelation contours appear to be self-similar, i.e., they share the same preferred orientation and the same aspect ratio. One also sees that $C_w(r,\tau;x)$ decays relatively slowly in the preference direction, but drops much faster in the direction that is perpendicular to the preference direction.

\begin{figure}
\center
\resizebox{1\columnwidth}{!}{%
  \includegraphics{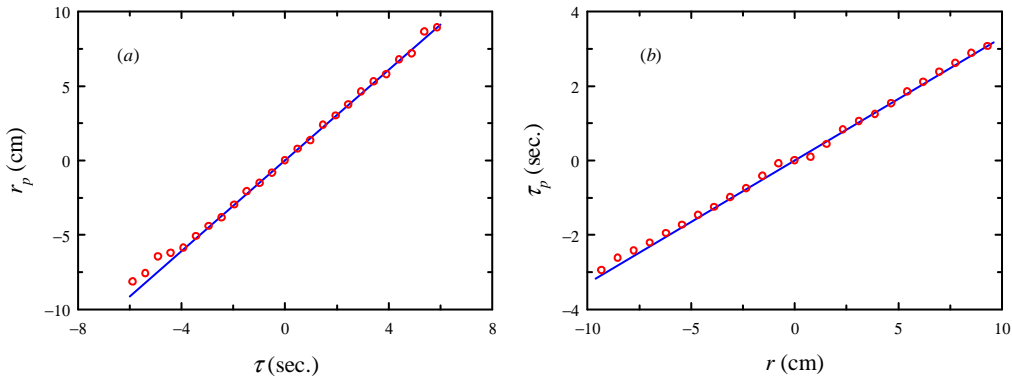}
} \caption{(\emph{a}) The measured peak position $r_p$ as a function of $\tau$. The solid line shows the fitted linear function, $r_p=U\tau$, with $U=1.52$ cm/s. (\emph{b}) The measured peak position $\tau_p$ as a function of $r$. The solid line shows the fitted linear function, $\tau_p=[U/(U^2+V^2)]r$, with $U/(U^2+V^2)=0.33$ s/cm. All data were obtained near the cell sidewall ($x=-21.73$ cm and $z=0$ cm).} \label{fig:fig4}
\end{figure}

To determine the characteristic velocities $U$ and $V$ in (\ref{eq:re}), note that from the conditions $\partial r_E/\partial r|_{\tau}=0$ and $\partial r_E/\partial \tau|_r=0$ we have, respectively,
\begin{equation}
\label{eqs:rp_tp}
r_p=U\tau \mbox{\ \ and\ \ } \tau_p=[U/(U^2+V^2)]r,
\end{equation}
where $r_p$ maximizes $C_w(r,\tau;x)$ for a given $\tau$ and $\tau_p$ is the peak position at which $C_w(r,\tau;x)$ reaches its maximum value for a fixed separation $r$ \cite[]{tong10pre}. Our results for the measured $r_p$ as a function of time separation $\tau$ are shown in figure \ref{fig:fig4}(\emph{a}). One sees clearly that $r_p$ increases with increasing $\tau$ because a longer time lag is needed to move velocity fluctuations across a larger separation. It is further seen that the increasing manner may indeed be described by a simple linear function, $r_p=U\tau$, with $U=1.52$ cm/s. How the measured $\tau_p$ varies with separation $r$ is shown in figure \ref{fig:fig4}(\emph{b}). Again, one sees that $\tau_p$ increases linearly with increasing $r$. A linear fit to the data yields $\tau_p=[U/(U^2+V^2)]r$ with $U/(U^2+V^2)=0.33$ s/cm. Note that the similar relation has also been observed for the scalar fields in the same system \cite[]{zhou08pre, tong10pre}. Taken together, we have $U=1.52$ cm/s and $V=1.51$ cm/s at $x=-21.73$ cm and $z=0$ cm. Comparisons between the fitted velocities $U$ and $V$ and the mean and the r.m.s. velocities of the flow will be presented in $\S$ \ref{sec:cell}.

\begin{figure}
\center
\resizebox{1\columnwidth}{!}{%
  \includegraphics{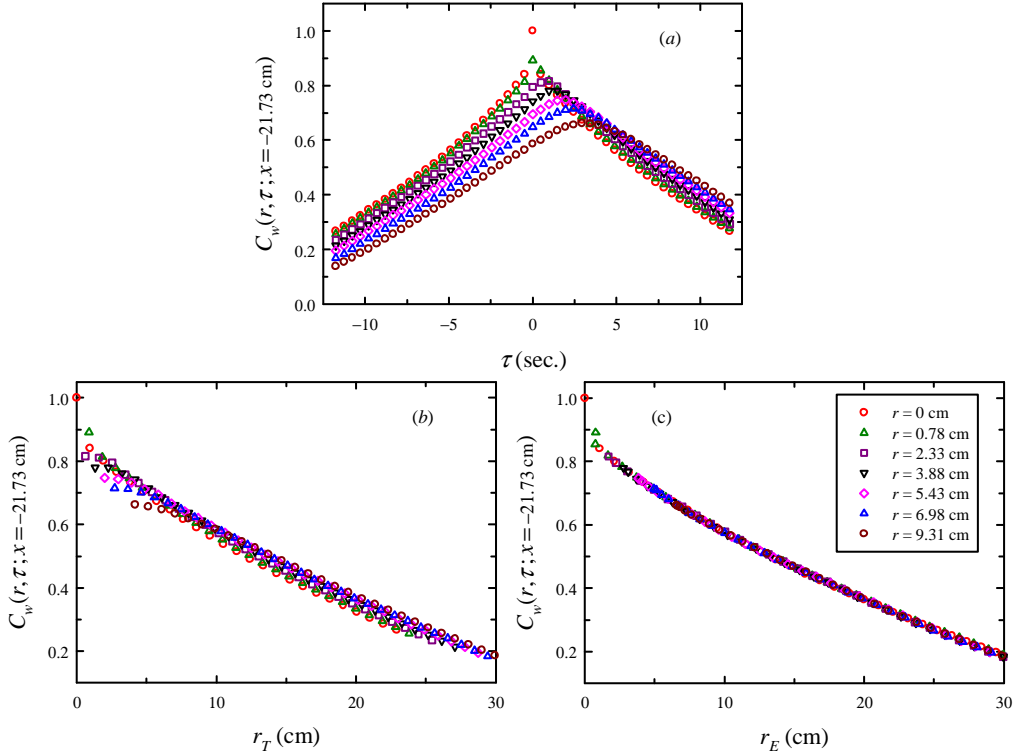}
} \caption{The space-time correlations $C_w(r,\tau; x)$ obtained near the cell sidewall ($x=-21.73$ cm and $z=0$ cm) for various values of $r=$0, 0.78, 2.33, 3.88, 5.43, 6.98, and 9.31 cm as functions of (\emph{a}) time separation $\tau$, (\emph{b}) the Taylor's separation $r_T=r-U_0\tau$ with the mean flow velocity $U_0=1.95$ cm/s, and (\emph{c}) the separation $r_E=\sqrt{(r-U\tau)^2+V^2\tau^2}$ with $U=1.52$ cm/s and $V=1.51$ cm/s.} \label{fig:fig5}
\end{figure}

With the obtained $U$ and $V$, we can now test the relation between $C_w(r,\tau;x)$ and $C_w(r_E,0;x)$. Figure \ref{fig:fig5}(\emph{a}) shows the evolution of $C_w(r,\tau;x)$ as a function of time separation $\tau$ for several different values of $r$. One sees that the measured $C_w(r,\tau;x)$ all have a single peak at the position $\tau_p$. The peak position $\tau_p$ increases with increasing $r$ [see also figure \ref{fig:fig4}(\emph{b})], meanwhile the correlation amplitude $C_w(r,\tau_p;x)$ decreases. This is because velocity fluctuations at two points decorrelate gradually when the separation between these two points increases. For comparison, we first plot in figure \ref{fig:fig5}(\emph{b}) the measured $C_w(r,\tau;x)$ as a function of the Taylor's separation $r_T$ with $r_T$ calculated from (\ref{eq:rt}) with the mean velocity $U_0=1.95$ cm/s, and then shows in figure \ref{fig:fig5}(\emph{b}) the measured $C_w(r,\tau;x)$ as a function of the separation $r_E$ with $r_E$ calculated from (\ref{eq:re}). In the figure, only positive parts of $C_w(r,\tau;x)$ are plotted because $C_w(r_E,0;x)$ is a symmetric function with respect to $r_E=0$. It is seen that when using the Taylor's hypothesis [see figure \ref{fig:fig5}(\emph{b})] the correlations could not collapse on top of each other, whereas, when using the elliptic model [see figure \ref{fig:fig5}(\emph{c})] the correlations all collapse well on top of each other, indicating that the space-time correlations $C_w(r,\tau;x)$ can be only determined by the space correlations $C_w(r_E,0;x)$ and the solution $r_E$ of (\ref{eq:re}).

\subsection{At the cell center}
\label{sec:center}

\begin{figure}
\center
\resizebox{1\columnwidth}{!}{%
  \includegraphics{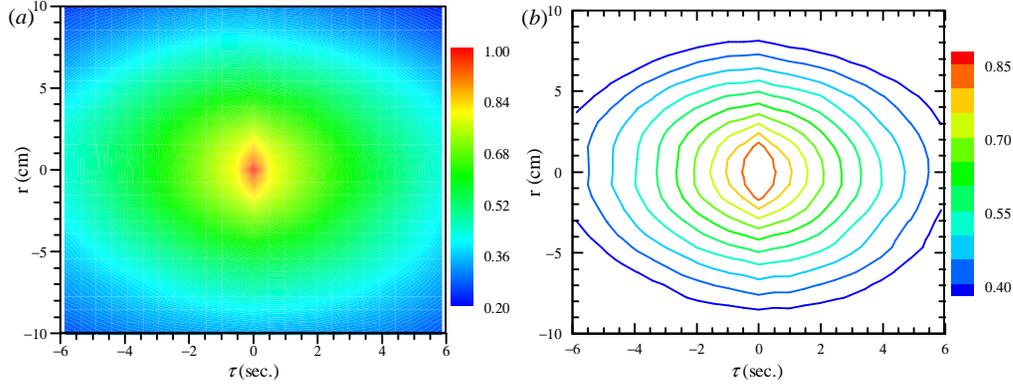}
} \caption{(\emph{a}) Space-time correlations $C_w(r,\tau; x)$ as a function of $r$ and $\tau$ measured at the cell center ($x=0$ cm and $z=0$ cm). Here, the amplitude of $C_w(r,\tau; x)$ is coded by color. (\emph{b}) The corresponding isocorrelation contours of $C_z(x,r,\tau)$ with the color-coded correlation amplitude varying from 0.4 to 0.85 at increments of 0.05 (outer to inner contours).} \label{fig:fig6}
\end{figure}

Let's now turn to the velocity field at the cell center ($x=0$ cm and $z=0$ cm), where the mean horizontal and vertical velocities are both nearly zero. As the velocity field in the cell's central region is approximately locally homogeneous and isotropic (Sun, Zhou $\&$ Xia 2006; Zhou, Sun $\&$ Xia 2008) and we also find that space-time correlations of the horizontal and vertical velocities share the same qualitative properties at the cell center, only the results of $C_w(r,\tau;x)$ will be presented in this subsection. Figure \ref{fig:fig6}(\emph{a}) shows the flood contours of space-time correlation $C_w(r,\tau; x)$ as a function of space separation $r$ and time separation $\tau$ and figure \ref{fig:fig6}(\emph{b}) shows the corresponding isocorrelation contours. Three features are worthy of note. (i) The measured $C_w(r,\tau;x)$ is a single-peak function with the peak locating at the origin and decays with increasing separations $r$ or $\tau$. (ii) All isocorrelation contours of $C_w(r,\tau;x)$ are closed curves and have an elliptic shape. (iii) The elliptic isocorrelation contours can be well described by a standard elliptic equation
\begin{equation}
\label{eq:seq}
\frac{\tau^2}{a^2}+\frac{r^2}{b^2}=1,
\end{equation}
i.e., the isocontours are aligned with the coordinate axis. Here, $a$ and $b$ are two parameters, determining the lengths of the major and minor axes of the standard ellipse.

\begin{figure}
\center
\resizebox{1\columnwidth}{!}{%
  \includegraphics{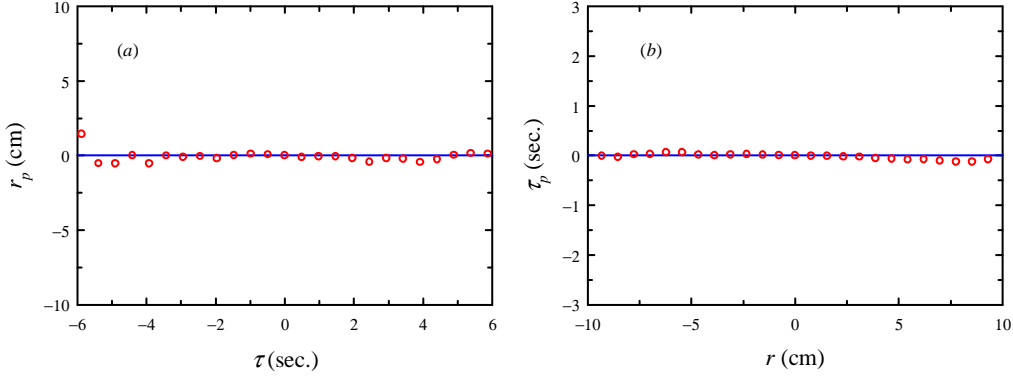}
} \caption{(\emph{a}) The measured peak position $r_p$ as a function of $\tau$. The solid line marks $r_p=0$. (\emph{b}) The measured peak position $\tau_p$ as a function of $r$. The solid line marks $\tau_p=0$. All data were obtained at the cell center ($x=0$ cm and $z=0$ cm).} \label{fig:fig7}
\end{figure}

\begin{figure}
\center
\resizebox{1\columnwidth}{!}{%
  \includegraphics{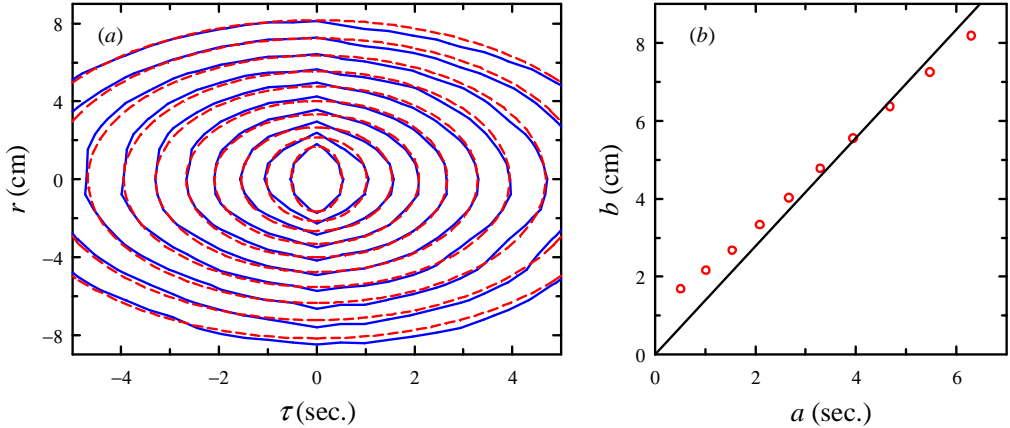}
} \caption{(\emph{a}) The blue solid curves are the isocorrelation contours of $C_z(x,r,\tau)$ measured at the cell center ($x=0$ cm and $z=0$ cm) with the correlation amplitude varying from 0.4 to 0.85 at increments of 0.05 (outer to inner contours). Here, we use the same date sets as figure \ref{fig:fig6}. The red dashed curves are the elliptic fittings of (\ref{eq:seq}) to the isocorrelation contours. (\emph{b}) The fitted parameters of the standard elliptic equation (\ref{eq:seq}): $b$ vs $a$. The solid line shows the best fit of $b=V a$, with $V=1.39$ cm/s. } \label{fig:fig8}
\end{figure}


Figure \ref{fig:fig7}(\emph{a}) shows the measured $r_p$ as a function of time separation $\tau$ and figure \ref{fig:fig7}(\emph{b}) shows the obtained $\tau_p$ as a function of space separation $r$. It is seen that $r_p$ and $\tau_p$ both vary around zero, implying $U\simeq0$ and $[U/(U^2+V^2)]\simeq0$ [see (\ref{eqs:rp_tp})]. To yield $V$, we note that when $U=0$, equation (\ref{eq:re}) can be rewritten as
\begin{equation}
r_E^2=r^2+V^2\tau^2.
\end{equation}
Comparing with the standard elliptic equation (\ref{eq:seq}), we have $b=V a$. Therefore, the value of $V$ can be estimated from the values of $a$ and $b$. In figure \ref{fig:fig8}(\emph{a}), we plot again the isocorrelation contours of $C_w(r,\tau;x)$ as the blue solid curves. In the figure, the best fittings of (\ref{eq:seq}) to the contours are plotted as the red dashed curves. It is seen that the fitted curves collapse well on top of the contours, further confirming that the isocorrelation contours have a shape of standard elliptic curves. Figure \ref{fig:fig8}(\emph{b}) shows the lengths of the $r$-axes of the fitted elliptic curves, $b$, as a function of the lengths of their $\tau$-axes, $a$. The solid line in the figure shows the best fit of $b=Va$ to the data, which gives $V=1.39$ cm/s. Taken together, we have $U\simeq0$ cm/s and $V=1.39$ cm/s at the cell center.

\begin{figure}
\center
\resizebox{1\columnwidth}{!}{%
  \includegraphics{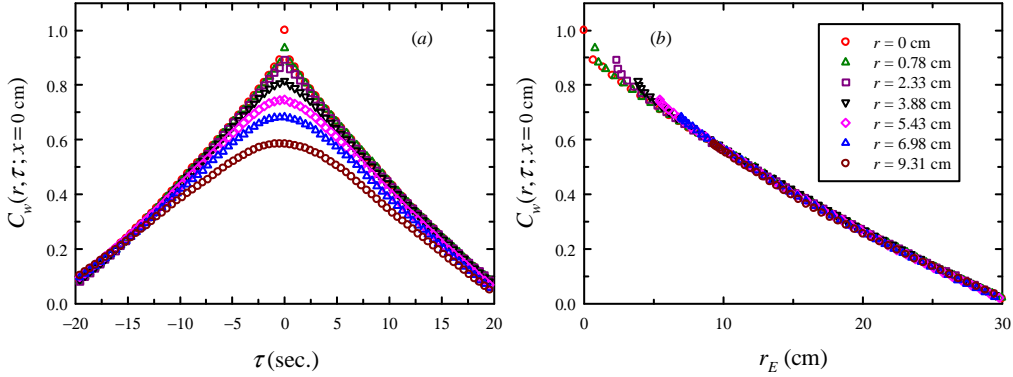}
} \caption{The space-time correlations $C_w(r,\tau; x)$ obtained at the cell center ($x=0$ cm and $z=0$ cm) for various values of $r=$0, 0.78, 2.33, 3.88, 5.43, 6.98, and 9.31 cm as functions of (\emph{a}) time separation $\tau$ and (\emph{b}) the separation $r_E=\sqrt{r^2+V^2\tau^2}$ with $V=1.39$ cm/s.} \label{fig:fig9}
\end{figure}

Figure \ref{fig:fig9}(\emph{a}) shows the space-time correlations $C_w(r,\tau; x)$ as a function of time separation $\tau$ for several different values of $r$. It is seen that $C_w(r,\tau; x)$ obtained at the cell center is also a single-peak function. However, unlike the case near the cell sidewall [see figure \ref{fig:fig5}(\emph{a})], the measured peak positions $\tau_p$ here do not vary with $r$, but all locate at positions around $\tau=0$ [see also figure \ref{fig:fig7}(\emph{b})]. This is because of the zero mean velocity at the cell center. Figure \ref{fig:fig9}(\emph{b}) shows the measured $C_w(r,\tau; x)$ as a function of $r_E$. One sees that when the solution $r_E$ of (\ref{eq:re}) is used, reasonable collapses among these correlations are achieved, further confirming the validity of the elliptic model for the present flow.

\subsection{Along the cell's horizontal and vertical central lines}
\label{sec:cell}

\begin{figure}
\center
\resizebox{1\columnwidth}{!}{%
  \includegraphics{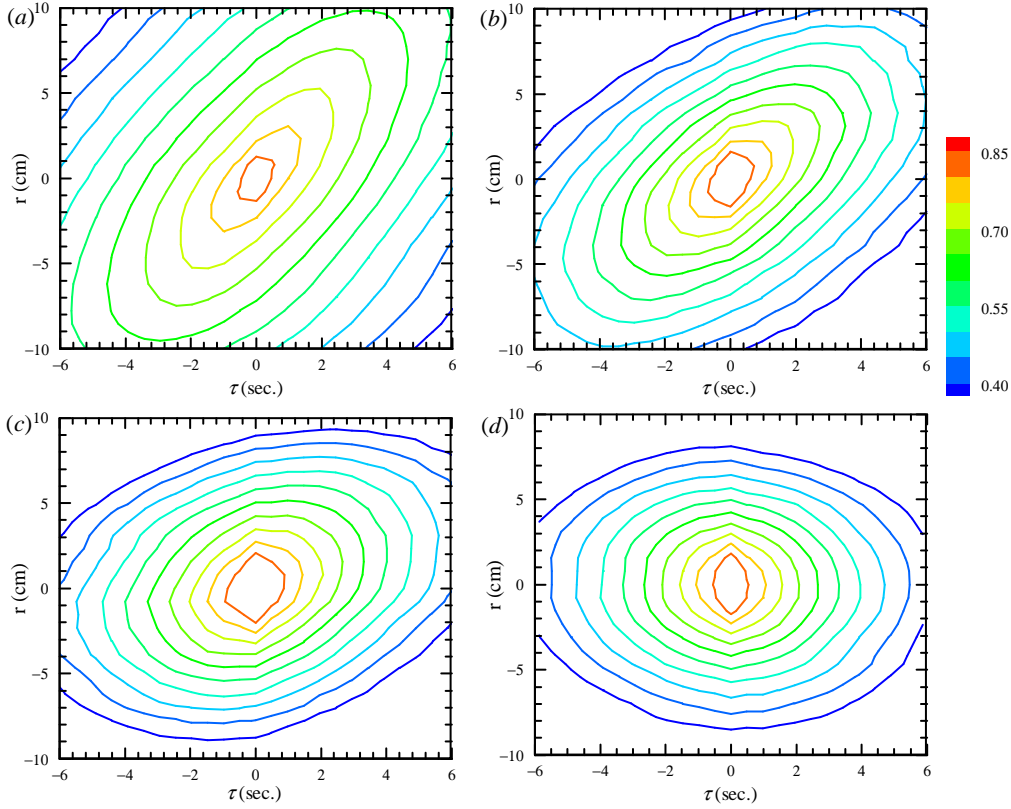}
}\caption{The isocorrelation contours of $C_w(r,\tau;x)$ as functions of time separation $\tau$ and space separation $r$ measured at the middle height of the cell ($z=0$ cm) at $x=-21.73$ (\emph{a}), $-13.97$ (\emph{b}), $-6.98$ (\emph{c}), and $0$ (\emph{d}) cm. The amplitude of the contours is coded by color and varies from 0.4 to 0.85 at increments of 0.05 (outer to inner contours). Note that (\emph{a}) is the same as figure \ref{fig:fig3}(\emph{b}) and (\emph{d}) is the same as figure \ref{fig:fig6}(\emph{b}). We replot them here for comparison.} \label{fig:fig10}
\end{figure}

Figures \ref{fig:fig10}(\emph{a})-(\emph{d}) show the evolution of the isocorrletion contours of the longitudinal space-time correlations $C_w(r,\tau;x)$ measured at the middle height of the cell ($z=0$ cm) at four different values of $x$ from near the cell sidewall to at the cell center. Figures \ref{fig:fig10}(\emph{a}) and (\emph{d}) are the same as figure \ref{fig:fig3}(\emph{b}) and figure \ref{fig:fig6}(\emph{b}), respectively. We replot these figures here for comparison. One sees that the contours are all elliptic closed curves with their preferred orientations. Furthermore, the slopes of the preferred orientations become smaller as the reference position movies from the wall towards the center. This is because the preferred orientations of the contours are directly related to the mean velocity of the flow \cite[]{he06pre, he09pre} and the mean vertical velocity $w_0$, after reaching its maximum value near the cell sidewall, decreases with the increasing distance from the wall \cite[]{qiu2001pre} (see also figure \ref{fig:fig1}).

\begin{figure}
\center
\resizebox{0.7\columnwidth}{!}{%
  \includegraphics{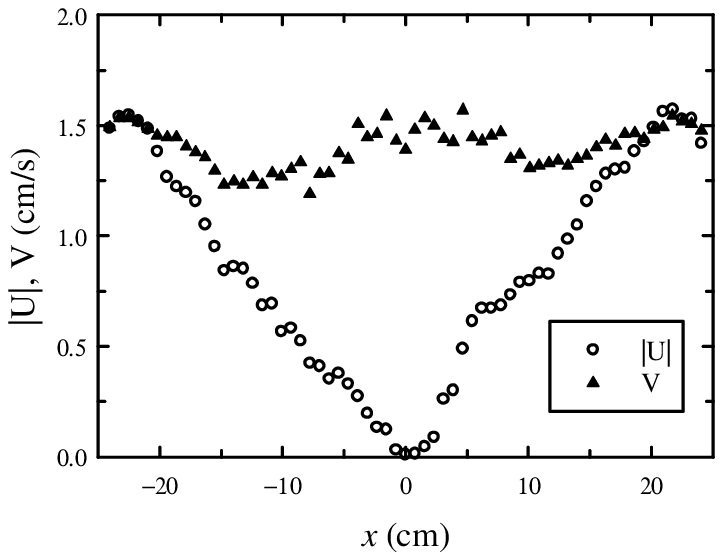}
} \caption{Comparison of the magnitudes of the characteristic velocities $U$ (circles) and $V$ (triangles) for $C_w(r,\tau;x)$ along the $x$-axis.} \label{fig:fig11}
\end{figure}

Direct comparison is made in figure \ref{fig:fig11} between the magnitudes of the characteristic velocities $U$ (circles) and $V$ (triangles) for $C_w(r,\tau;x)$ along the $x$-axis. It is seen that $|U|\simeq0$ at the cell center and increases along the cell's diameter at the middle height of the cell from the cell center to the sidewall and the maximization of $|U|$ occurs near the cell sidewall, while the variation of $V$ is much weaker. Note that the validity of Taylor's frozen-flow hypothesis requires $|U|\gg V$. However, the figure shows clearly that $|U|\gg V$ is not the case, i.e., $|U|$ and $V$ are approximately the same near the sidewall, while at the cell center the value of $V$ is even much larger than that of $|U|$. This further conforms that Taylor hypothesis does not hold in the present system.

\begin{figure}
\center
\resizebox{1\columnwidth}{!}{%
  \includegraphics{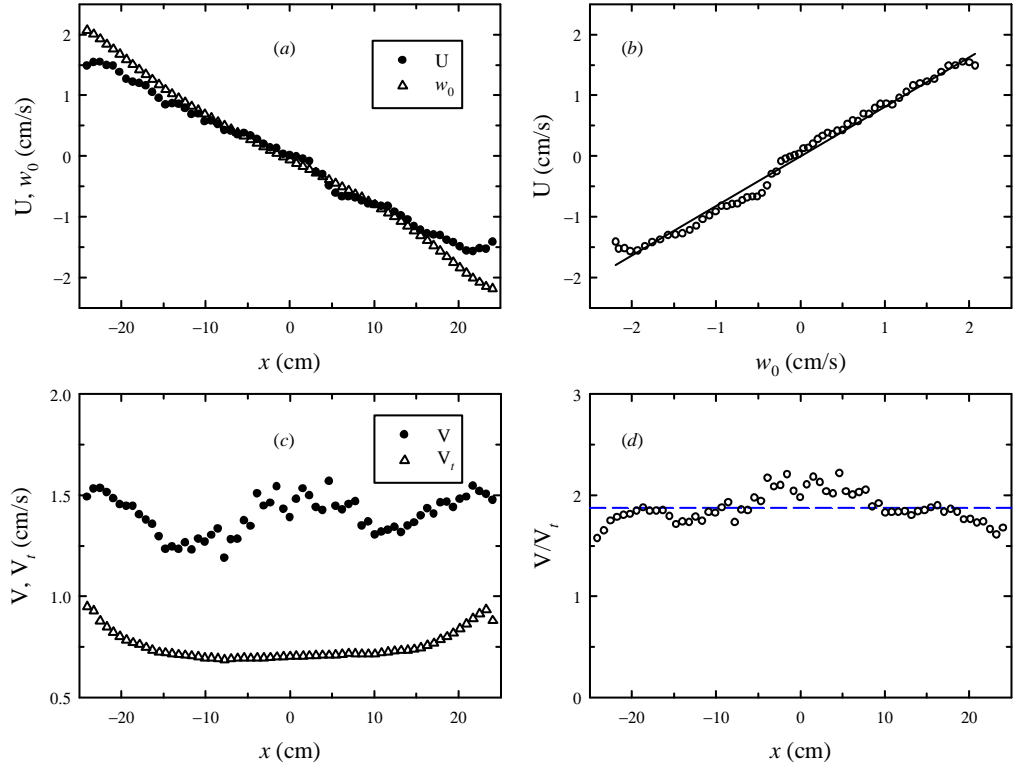}
} \caption{(\emph{a}) Comparison of the characteristic velocity $U$ (solid circles) for $C_w(r,\tau;x)$ and the mean velocity of the flow $w_0$ (open circles) along the $x$-axis. (\emph{b}) $U$ vs $w_0$. The solid line shows the linear fit to the data, $U=0.82w_0$. (\emph{c}) Comparison of the characteristic velocity $V$ for $C_w(r,\tau;x)$ (solid circles) and the theoretical prediction $V_t$ (open circles) of the elliptic model along the $x$-axis. (\emph{d}) $V/V_t$ vs $x$. The solid line marks the mean value 1.87 of the ratios.} \label{fig:fig12}
\end{figure}

To further test the elliptic model, we compare in figure \ref{fig:fig12} the measured values of the characteristic velocities $U$ and $V$ and their theoretical predictions $U_t$ and $V_t$. Based on Navier-Stokes equation, \cite{he09pre} showed that $U_t$ is a characteristic convection velocity proportional to the mean velocity of the flow and $V_t$ is the sum of the random sweeping velocity and the shear-induced velocity, i.e. $V_t=\sqrt{S^2\lambda^2+w^2_{rms}}$, where the subscript "t'' indicates theoretical predications and $S$ and $\lambda$ are, respectively, the shear rate and the Taylor microscale of the flow. For the vertical velocity profile along the $x$-axis \cite[]{qiu2001pre,sun2005pre}, the shear rate was evaluated as $S\simeq2(W_0)_{max}/D$ \cite[]{tong10pre}, where $(W_0)_{max}$ is the maximal value of the vertical velocity along the $x$-axis. The Taylor microscale $\lambda(x)$ was estimated using the equation $C_w(r_E,0;x)\simeq1-(r/\lambda(x))^2$ for $|r|<1$ cm, where the space autocorrelation function $C_w(r_E,0;x)$ was obtained from the time autocorrelation function $C_w(0,\tau;x)$ using the relation (\ref{eq:re}) with $r=0$. We note that $S\lambda(x)\ll w_{rms}(x)$ for all measuring positions. Comparison is made in figure \ref{fig:fig12}(\emph{a}) between $U$ (solid circles) and the mean vertical velocity $w_0$ (open circles). One sees that both $U$ and $w_0$ decrease with increasing $x$, but the magnitude of $U$ seems to be systematically smaller than that of $w_0$. Figure \ref{fig:fig12}(\emph{b}) shows the measured $U$ as a function of $w_0$. It is seen that $U$ increases with increasing $w_0$ and the increasing manner may be described by a simple linear function $U=0.82w_0$. The solid line in figure \ref{fig:fig12}(\emph{b}) shows the fitting linear function. Figure \ref{fig:fig12}(\emph{c}) shows the comparison of $V$ and $V_t$. One sees that the experimentally measured $V$ is larger than its theoretical predictions $V_t$ for all measuring positions. The ratio of $V$ to $V_t$ along the $x$-axis is plotted in figure \ref{fig:fig12}(\emph{d}). In the figure, the dashed line marks the mean value 1.87 of the ratios. One sees that all data points vary around the dashed line. This seems to suggest a constant ratio between $V$ and $V_t$, i.e., the experimentally measured $V$ is proportional to its theoretical predications $V_t$.

\begin{figure}
\center
\resizebox{1\columnwidth}{!}{%
  \includegraphics{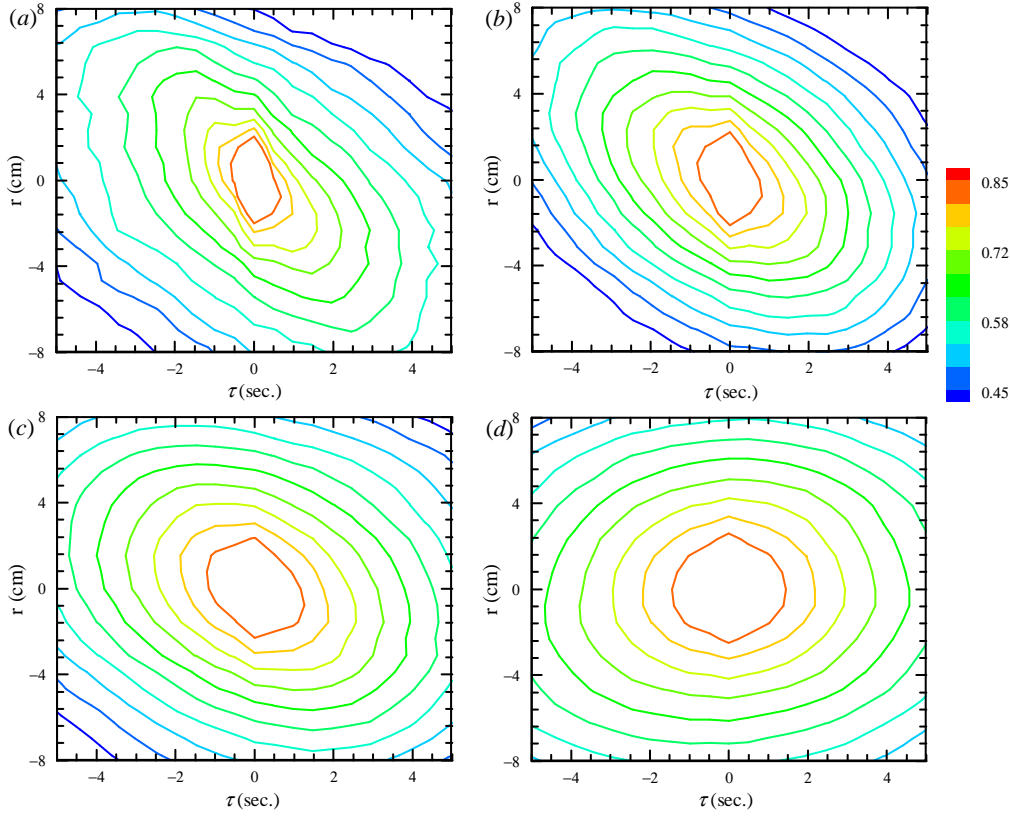}
}\caption{The isocorrelation contours of $C_u(r,\tau;z)$ as functions of time separation $\tau$ and space separation $r$ measured along the cell's central vertical axis ($x=0$ cm) at $z=-21.73$ (\emph{a}), $-13.97$ (\emph{b}), $-6.98$ (\emph{c}), and $0$ (\emph{d}) cm. The amplitude of the contours is coded by color and varies from 0.4 to 0.85 at increments of 0.05 (outer to inner contours).} \label{fig:fig13}
\end{figure}

\begin{figure}
\center
\resizebox{0.7\columnwidth}{!}{%
  \includegraphics{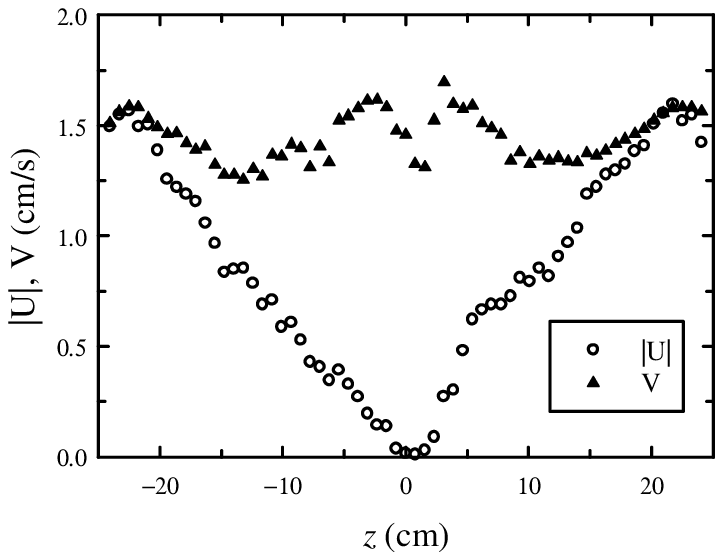}
} \caption{Comparison of the magnitudes of the characteristic velocities $U$ (circles) and $V$ (triangles) for $C_u(r,\tau;z)$ along the $z$-axis.} \label{fig:fig14}
\end{figure}

\begin{figure}
\center
\resizebox{1\columnwidth}{!}{%
  \includegraphics{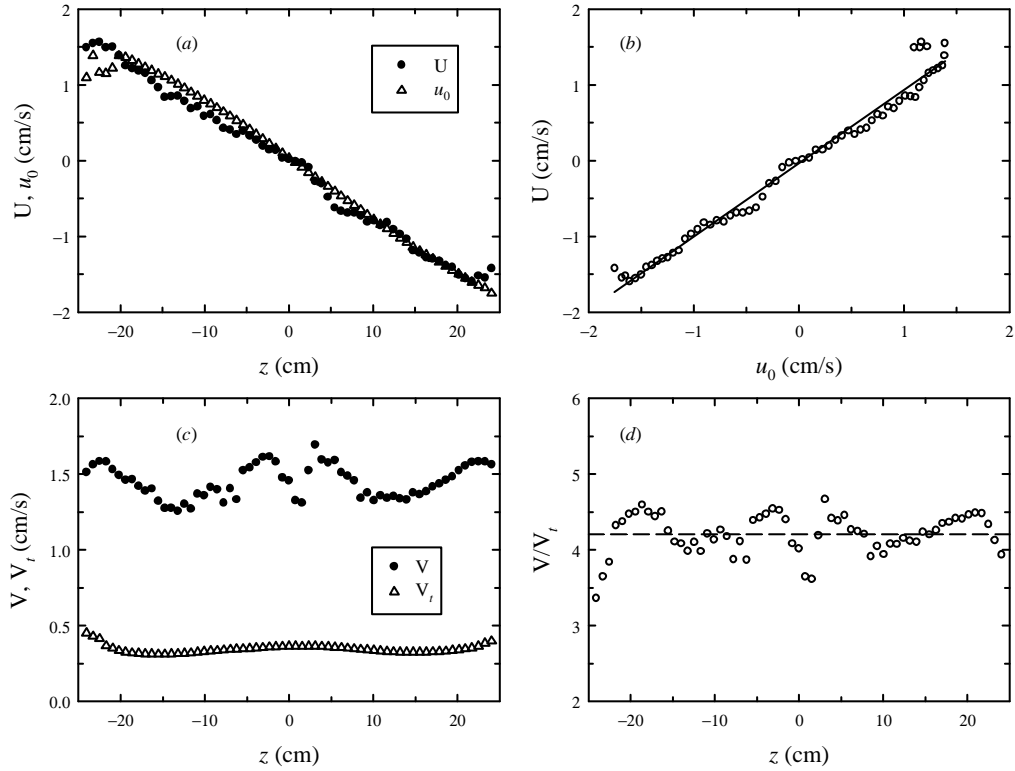}
} \caption{(\emph{a}) Comparison of the characteristic velocity $U$ (solid circles) for $C_u(r,\tau;z)$ and the mean velocity of the flow $u_0$ (open circles) along the $z$-axis. (\emph{b}) $U$ vs $u_0$. The solid line shows the linear fit to the data, $U=0.97u_0$. (\emph{c}) Comparison of the characteristic velocity $V$ for $C_u(r,\tau;z)$ (solid circles) and the theoretical prediction $V_t$ (open circles) of the elliptic model along the $z$-axis. (\emph{d}) $V/V_t$ vs $z$. The solid line marks the mean value 4.21 of the ratios.} \label{fig:fig15}
\end{figure}

Finally, we study the longitudinal space-time correlations $C_u(r,\tau;z)$ for the horizontal velocity along the cell's central vertical axis. Figures \ref{fig:fig13}(\emph{a})-(\emph{d}) show the evolution of the isocorrelation contours of $C_u(r,\tau;z)$ obtained at $x=0$ and four different values of $z$. Again, one sees that all contours are elliptic closed curves with their preferred orientations. The preferred orientations guide along the second and fourth quadrants, which is due to the negative mean velocities at the measuring positions (see figure \ref{fig:fig1}), and the slopes of the preferred orientations decrease with the increasing distance from the plate, which is due to the decrease of the magnitude of the mean horizontal velocity (see figure \ref{fig:fig1}). Figure \ref{fig:fig14} shows the comparison of $|U|$ and $V$. It is also seen that $|U|$ and $V$ are nearly the same near the top and bottom plates, while $V$ is much larger than $|U|$ at the cell center. The fact that $|U|\gg V$ does not hold for the horizontal velocity again indicates the invalidity of Taylor's frozen-flow hypothesis in the present flow.

Figure \ref{fig:fig15}(\emph{a}) shows the comparison of the measured $U$ and the mean horizontal velocity $u_0$ along the cell's central vertical axis. It is seen that $U$ and $u_0$ are approximately the same for all values of $z$. Figure \ref{fig:fig15}(\emph{b}) shows $U$ as a function of $u_0$. The best linear fit to the data yields $U=0.97u_0$, again indicating that $U$ is proportional to $w_0$. Direct comparison is made in figure \ref{fig:fig15}(\emph{c}) between $V$ and $V_t$ ($=\sqrt{S^2\lambda^2+u^2_{rms}}$). Here, the shear rate was estimated as $S\simeq2(U_0)_{max}/H$ with $(U_0)_{max}$ being the maximal horizontal velocity along the $z$-axis and the Taylor microscale $\lambda(z)$ was evaluated from the space autocorrelation function $C_u(r_E,0;z)$. Again, we find that $S\lambda(z)\ll u_{rms}(z)$ for all measuring positions. In figure \ref{fig:fig15}(\emph{c}), one sees that similar to the case of $C_w(r,\tau;x)$, the values of $V$ are also larger than those of $V_t$. Nevertheless, the ratio $V/V_t$ shown in figure \ref{fig:fig15}(\emph{d}) varies around its mean value 4.21, suggesting that $V$ is proportional to $V_t$. Taken together, our results reveal that $C_u(r,\tau;z)$ shares the same qualitative properties as $C_w(r,\tau;x)$.

\section{Conclusion}

To conclude, we have presented an systematic experimental study of the velocity field in a cylindrical turbulent Rayleigh-B\'{e}nard (RB) convection cell with unit aspect ratio using water as working fluid. The two-dimensional velocity field in the vertical circulation plane of the large-scale circulation was measured via the particle image velocimetry (PIV) technique and the longitudinal space-time cross-correlation functions for both the horizontal and vertical velocities, $C_u(r,\tau;z)$ and $C_w(r,\tau;x)$, were investigated in great detail. Our results show that the isocorrelation contours of space-time correlations are elliptic closed curves and the space-time correlations $C_u(r,\tau;z)$ and $C_w(r,\tau;x)$ can be related to the space correlations $C_u(r_E,0;z)$ and $C_w(r_E,0;x)$, respectively, via the elliptic relation (\ref{eq:re}), i.e. $r_E^2=(r-U\tau)^2+V^2\tau^2$. The characteristic velocities $U$ and $V$ were then calculated and studied. We find that the magnitude of $U$ reaches its maximum value near the sidewall and plates and decreases when away from the walls and plates, while the position-dependence of $V$ is much weaker. Specifically, at the cell center we have $U\simeq0$ and hence the relation (\ref{eq:re}) becomes $r_E^2=r^2+V^2\tau^2$ with its major and minor axes coinciding with the $\tau$- and $r$-axis. Note that such relation has the same form as Kraichnan's sweeping-velocity hypothesis \cite[]{kraichnan64}. Direct comparison of the values of $U$ and $V$ and their theoretical predictions further show that $U$ is proportional to the mean velocity of the flow, while $V$ is systematically larger than its prediction.

Our results validate the elliptic model in turbulent RB convection, where Taylor's frozen-flow hypothesis does not hold due to the relatively large values of the r.m.s. velocity. As pointed out by \cite{he06pre}, the elliptic model is developed based on a second-order approximation, while Taylor's hypothesis implies a first-order approximation. Thus, the elliptic model is a generation of Taylor's frozen-flow hypothesis and Taylor's hypothesis is only a special case of the elliptic model [i.e., when $V=0$ the elliptic model's relation (\ref{eq:re}) is degenerated to the Taylor's relation (\ref{eq:rt})]. Like Taylor's hypothesis, the elliptic model could also be used to translate time series to space series. This is because the correlation function is a basic quantity and most statistical properties interested in the field of turbulence, such as structure function and power spectrum, can be obtained theoretically from the correlation functions. In fact, previous work by \cite{tong10pre} has used the model to translate the temperature power spectrum from time domain to space domain.

An important implication of the elliptic model is that when $r=0$, the relation (\ref{eq:re}) becomes
\begin{equation}
r_E=(U^2+V^2)^{1/2}\tau.
\label{eq:re0}
\end{equation}
In this case, $r$ is still proportional to $\tau$ but just that the proportionality constant is $(U^2+V^2)^{1/2}$, rather than the mean velocity $U_0$ as stated in the Taylor's hypothesis [see (\ref{eq:rt}) with $r=0$]. This implies that if one is ONLY interested in the scaling exponents of $C(r,0)$ in space domain, such scaling properties can still be correctly obtained by studying the scaling of $C(0,\tau)$ in time domain in the elliptic model even though Taylor's hypothesis is not valid. We note that Taylor's hypothesis has been widely used to study the scaling behaviors of structure functions and power spectrum of the velocity and temperature fields in turbulent RB convection \cite[]{lx10arfm}. However, it has long been known in the field that the condition for Taylor's hypothesis is not often net \cite[]{lx10arfm}, and hence the results based on Taylor's hypothesis is questionable and not convincible. Here, the relation (\ref{eq:re0}) implies that one does not really need the validity of Taylor's hypothesis to reconstruct the space series from the measured time series. However, Taylor's hypothesis could not yield the correct scaling ranges in space domain, which could only be obtained upon the transform of (\ref{eq:re0}) in the elliptic model.

It should be noted that in the original elliptic model \cite[]{he06pre} the correlations $R(r,\tau)=\langle v(\textbf{\emph{x}}+\textbf{\emph{r}},t+\tau)v(\textbf{\emph{x}},t)\rangle$ were considered, while we studied the normalized quantities $C(r,\tau)=R(r,\tau)/v_{rms}(\textbf{\emph{x}})v_{rms}(\textbf{\emph{x}}+\textbf{\emph{r}})$ here. The elliptic relation (\ref{eq:re}) holds for both $R(r,\tau)$ and $C(r,\tau)$. However, if the r.m.s. velocity $v_{rms}(\textbf{\emph{x}})$ depends on position, the characteristic velocities $U$ and $V$ for $R(r,\tau)$ may be different from those for $C(r,\tau)$. We note that previous works usually focused on $C(r,\tau)$ \cite[see, e.g.,][]{tong10pre}. In the present work, we have also checked the properties of $R(r,\tau)$ and the similar results were obtained.



\begin{acknowledgments}
The experiments were carried out in the Chinese University of Hong Kong. We gratefully acknowledge Prof. Ke-Qing Xia for making the PIV data available to us. We thank also Prof. Guo-Wei He and Dr. Xiao-Zhou He for helpful discussions. This work was supported by Natural Science Foundation of China (Nos. 11002085, 11072139), ``Pu Jiang" project of Shanghai (No. 10PJ1404000), ``Chen Guang" Project of Shanghai (No. 09CG41), E-Institutes of Shanghai Municipal Education Commission, and Shanghai Program for Innovative Research Team in Universities.
\end{acknowledgments}

\bibliographystyle{jfm}

\end{document}